\documentclass[pre, twocolumn, amsmath, amssymb, showkeys]{revtex4}

\usepackage{graphicx}

\begin{document}

\title{Speculations on the emergence of self-awareness in big-brained organisms}

\author{Emmanuel Tannenbaum}
\email{emanuelt@bgu.ac.il}
\affiliation{Ben-Gurion University of the Negev,
Be'er-Sheva, Israel}

\begin{abstract}

This paper argues that self-awareness is a learned behavior that emerges in organisms whose brains have a sufficiently integrated, complex ability for associative learning and memory.  Continual sensory input of information related to the organism causes the organism's brain to learn the physical characteristics of the organism, in the sense that neural pathways are produced that are reinforced by, and therefore recognize, various features associated with the organism.  More precisely, continual sensory input leads to the formation of a set of associations that may be termed an organismal ``self-image".  The formation of a self-image, combined with an ability for sufficiently complex associative memory and learning, provides a mechanistic basis for the emergence of various behaviors that are typically associated with self-awareness.  Self-recognition is the classic example.  However, self-awareness in humans includes additional behaviors such as recognition of self-awareness, the concept of ``I", and various existential and religious questions.  After providing the basic mechanistic basis for the emergence of an organismal self-image, this paper proceeds to go through a representative list of behaviors associated with self-awareness, and shows how associative memory and learning, combined with an organismal self-image and, in the case of humans, language, leads to the emergence of these various behaviors.  This paper also discusses various tautologies that invariably emerge when discussing self-awareness, that ultimately prevent an unambiguous resolution to the various existential issues that arise.  We continue with various speculations on manipulating self-awareness, and discuss how concepts from set and logic may provide a highly useful set of tools in computational neuroscience for understanding the emergence of higher cognitive functions in complex organisms.  The existence of other types of awareness, and the role of mirror neurons in the emergence of self-awareness, are also briefly discussed. 

\end{abstract}


\maketitle

\section{Introduction}

Self-awareness is one of the most mysterious phenomena in the natural world.  Inanimate matter, through the process of replicative selection, has managed to produce biochemical machines that can recognize themselves.  The existence of self-awareness has provoked much philosophical debate over the centuries, and, within a theological context, is intimately connected with the existence of a metaphysical ``soul" \cite{SELFAWARE1}.

While it was once widely believed that only humans were truly self-aware, it was later discovered that the great apes could recognize their own reflections, and therefore had a much higher level of self-awareness than previously thought.  It was later discovered that dolphins also exhibit this level of self-awareness.  Recently, elephants have been added to the list of animals capable of recognizing their own reflections \cite{APE1, DOLPHIN1, ELEPHANT1}.

A major difficulty in understanding self-awareness has been its theological association with the concept of a ``soul".  Because the phenomenon was regarded as mysterious, there have been relatively few scientific attempts to attack the problem.  It was essentially viewed as an epiphenomenon associated with a sufficiently developed intelligence \cite{EPIPHENOMENON1, EPIPHENOMENON2}.

The development of sophisticated brain imaging techniques has allowed researchers to locate the regions of the brain where self-awareness is located \cite{SELFAWARE2}.  Specifically, it is believed that self-awareness is processed in a region of the brain called the Superior Frontal Gyrus 
\cite{SELFAWARE2}.  However, the origins of such brain structures are essentially a mystery.  That is, while there have been speculations as to the evolutionary pressures that would select for self-awareness \cite{SELECTION1, SELECTION2, SELECTION3}, it is not clear whether self-awareness was independently ``hard-wired" into the brain, or whether it is a property that emerges in a sufficiently developed central nervous system.  If it is a so-called ``emergent property," then it is not clear what are the underlying dynamics of a highly developed brain that leads to self-aware behavior.

A mechanistic study of self-awareness is possible, if we abandon traditional philosophical attempts to define what it is (which is a meaningless question, in any event), and instead attempt to study its operational aspects.  According to Turing, a mechanistic approach to self-awareness requires us to determine what behaviors we would associate with self-awareness, so that self-awareness, in a precise sense, becomes defined by these behaviors \cite{TURINGTEST}.  The next step is to then determine what is the mechanistic basis for the emergence of such behaviors.  Finally, we should then seek to determine how such behaviors emerge in organismal brains.  That is, are they hard-wired, or are they learned?  

Here, we argue that self-awareness is essentially a learned behavior, that emerges in organisms with a sufficiently complex and highly integrated ability for associative memory and learning.  The idea is that the organism's brain perceives the external world from a specific vantage point, and receives a continual set of inputs related to the organism as a result of this vantage point.  This continual set of inputs related to the organism leads to the formation of neural pathways that define a set of associations related to various aspects of the organism, resulting in behavior consistent with self-recognition.  We call this set of associations the organismal ``self-image". 

This paper is organized as follows:  In the following section (Section II), we describe in further detail our speculation regarding the emergence of self-awareness.  In Section III, we consider various aspects of self-awareness, and speculate as to the possible mechanistic basis for their emergence.  Specifically, we consider self-recognition, the emergence of the concept of ``I/Me", and the emergence of an organism's awareness of being self-aware.  This last phenomenon in particular has been the subject of intense philosophical and theological debate over the centuries \cite{SELFAWARE1}.  In Section IV, we consider aspects of self-awareness that are apparently unique to humans, namely various existential and religious concepts such as the mind and soul, solipsism, the idea that reality is an illusion, and various conceptions of divinity.  We also discuss the problem of tautologies that arises when considering the various existential questions associated with self-awareness, a problem that prevents an unambiguous resolution of the problem of self-awareness and the various associated existential issues.  In Section V, we discuss various experiments that could be used to manipulate self-awareness, as well as test some of the hypotheses presented here.  Furthermore, we speculate on how certain kinds of metaphysical constructs, such as mind-body separation and the delocalization of mind in several bodies, may in principle be physically realizable.  We conclude with a brief discussion of a number of additional issues related to self-awareness and computational neuroscience, such as the role of mirror neurons, other forms of awareness, the concept of qualia, and the role that set and logic theory could play in understanding various neural structures associated with higher cognitive functions in complex organisms.  

\section{Self-Awareness as a Learned Behavior}

All sufficiently complex terrestrial life has a brain which processes the sensory input obtained by the 
organism's sensory organs.  The five senses familiar to humans are sight, hearing, touch, taste, and
smell.  In addition to these basic senses, which provide the organismal brain with external information about the world, neural connections throughout the body provide the brain with information about various parts of the body in which it sits \cite{ASTANNEN}.    

A seemingly obvious point to note is that the organism's brain sits inside the organism's head, and that the organism's various sensory organs and internal sensory connections are located in specific parts of the body, which are wired to the organism's brain in a specific way.  Therefore, the organism's brain is obtaining information about the world from a specific vantage point.  This vantage point is such that the flow of information from the environment always contains a constant subset of information related to the organism.  Using humans as an example, the human brain constantly receives visual information from the human in which it sits.  This visual information consists of various parts of the human organism's body, such as hands, arms, tip of the nose, feet, legs, torso, etc.  It should be emphasized that, from the perspective of the brain, the environmental inputs also include the ``internal'' sensory inputs, since these provide information about certain aspects of the material world, in this case coming from the organism itself.

Since this subset of ``self-information" is constant, the organism's brain can re-wire itself in response to this information flow, until neural pathways are formed that respond to stimuli connected to the organism.  
The end result of this ``self-learning" process is the formation of neural pathways that respond to inputs related to the organism, and therefore recognize the organism.  The neural pathways are said to then store a kind of ``self-image" that is associated with the organism.  Inputs that are related to this self-image then trigger the relevant neural pathways, resulting in self-recognition.  Given that it is believed that neural pathway formation is driven by a reward-punishment-based selection process, we argue that the emergence of an organismal ``self-image" should be a self-reinforcing process that automatically extracts the subset of information flow associated with the organism itself.  

It makes sense that self-awareness should be processed in regions of the brain that integrate many different sources of input together.  First of all, such a region of the brain will have a stronger total signal strength connected to the organism.  Furthermore, since the different sources of external inputs carry different aspects of information about the organism, these respective aspects will all be strongly correlated with one another.  In an organism with a sufficiently powerful associative memory and learning ability, these various self-inputs will result in the formation of a highly correlated, multi-dimensional ``super-input."  Because sources of input that derive from the environment, and not the organism, are constantly changing, the strength of the multi-dimensional signal obtained from the organism becomes proportionally stronger as the number of inputs that are integrated together increases (in a rough analogy, the environmental inputs are like noise terms that are averaged away as more and more sources of input are integrated together).  

It should be emphasized that the self-image is not necessarily a well-defined set of neural pathways that represents a specific internal picture of the organism.  Rather, it should be thought more of as a set of associations generated by the continual input of information related to the organism itself.  As will be seen later, these associations between various aspects of the organism provide a mechanism for the organism to learn certain behaviors that are associated with what is termed self-awareness.  In this vein, it should be noted that self-awareness has previously been referred to as a ``structure with variations" \cite{ASTANNEN, STRUCTVARY}.  

Furthermore, while these associations are instantiated by various neural connections, they do not consist exclusively of associations of pathways that respond to sensory inputs related to the organism.
While such associations may provide a nucleus of pathways that form the initial self-image, the formation of memories involving the self-image may themselves become part of the self-image, so that the number of associations defining the self-image can increase in time (though presumably, the size of the self-image saturates, since the brain contains a finite number of neurons).  

As an example, while an adult human organism looks differently than it did as a child, it may nevertheless associate a childhood picture with itself.  This could conceivably occur in a variety of ways, which are not necessarily mutually exclusive:  (1)  The picture is associated with certain childhood memories, which were formed with the organismal brain perceiving the world from a certain vantage point (due to the specific way that the brain receives sensory inputs from the environment).  This vantage point is most closely associated with the vantage point of the organismal brain in the adult body, which is itself associated with the organismal self-image.  (2)  The picture is associated with certain childhood memories, which are associated with certain aspects of the child organism that are identical in the adult organism, which are then associated with the organismal self-image.  An example is an organismal label such as a name.

Therefore, we argue that memory is crucial for the emergence of self-awareness, since it allows for the formation of neural pathways that can become associated with the self-image, and therefore become part of the self-image itself.  Presumably, then, the larger the memory capacity of an organismal brain, the larger and more complex the self-image that can be formed, and hence the ``higher" the level of self-awareness that can be achieved.

The arguments presented here suggest that self-awareness should emerge in slowly replicating organisms.  For slowly replicating organisms, there is a strong selection pressure to adopt a big-brained survival strategy, in order to ensure organismal survival to allow for reproduction.  Large brained organisms are likely under selection pressures to evolve a brain topology that allows for the integration of many different sources of input together.  This ensures that the brain operates as a unit, and not as several independent modules that can possibly engage in contradictory behaviors and thereby adversely affect fitness.  In addition, an integrated brain topology allows for the co-processing of many different sources of information, which presumably leads to an improved ability to choose an optimal survival strategy.

The emergence of an integrated neural topology therefore parallels integration processes seen in the emergence of other complex systems, such as multicellular organisms and networked societies.  In general, systems that evolve collective replicative strategies will be under selection pressures to integrate and coordinate the actions of their various component parts, since a system that behaves as a single entity will likely have a significant survival advantage over systems in which the components function independently and do not communicate with one another.

It should be emphasized that the idea of external and internal sensory inputs driving the emergence of self-awareness has been previously considered by a number of authors \cite{ASTANNEN, OREGAN}.  Indeed, we believe that many of the arguments that are made in this paper have been already presented in \cite{ASTANNEN}.  Nevertheless, we believe that this paper differs from previous works on the subject in that it specifically identifies associative learning as the underlying mechanistic basis for the emergence of the various behaviors defining self-awareness.  Therefore, the discussion in this paper might provide a more useful framework for understanding the evolution of self-awareness, and for facilitating the construction of machines that can exhibit self-aware behaviors.

\section{The Emergence of Three ``Classic" Aspects of Self-Awareness}

In this section, we speculate on the emergence of three ``classic" aspects of self-awareness:  (1) Self-recognition in the mirror.  (2)  The concept of ``I" or ``Me".  (3)  The phenomenon whereby an organism is aware of being self-aware.

\subsection{Self-recognition in the mirror}

One of the standard tests for determining whether an organism is self-aware is to see if it recognizes itself when placed in front of a mirror.  Thus far, only humans, the great apes, dolphins, and, more recently, elephants, have been shown to exhibit self-recognition.  Other animals simply ignore the image, or see it as a rival organism \cite{APE1, DOLPHIN1, ELEPHANT1}.

If an organism has a brain that is capable of forming a self-image, as described above, then, when the organism sees its reflection in the mirror, various aspects of that reflection will trigger stimulation of
the neural pathways containing the self-image.  The result is that the organism recognizes the reflection in the mirror, and associates it with the various self-inputs that produced the self-image in the first place.
Therefore, if, as is done in a number of self-recognition experiments, an ``X" is painted on the organism's forehead \cite{ELEPHANT1}, a sufficiently intelligent organism will understand that the image in the mirror will be seen touching the ``X" if the organism touches the region of the body that the location of the ``X" most closely corresponds to in the self-image, i.e. the forehead (this presumably supposes that the organism has an idea of what a mirror does, otherwise the organism might try to touch the ``X" of the reflection). 

While the paragraph above provides a general framework for understanding the phenomenon of self-recognition, it is still useful to discuss in further detail the various ways by which the organismal brain can associate the image in the mirror with the organismal self-image, and thereby make the reflection a part of the self-image.

If the organism moves a certain body part, say an arm, in front of the mirror, it may notice a correlation between the motion of the reflected body part and its own body part.  The result is that the reflected body part becomes associated with the actual body part.  Furthermore, the organism may notice that the reflected body part is connected to the body of another organism, so that the moving body part in the reflection is associated with the entire body in the reflection.  If the organism has an ability to perceive spatial relations, then, since the reflected body part may be mapped onto the actual body part with a certain transformation, the organismal brain might instinctively apply the same transformation to the rest of the body in the image, which would result in the body in the reflection becoming associated with the organism's actual body.  The end result of this is that the reflection in the mirror becomes associated with the organism's self-image.  Indeed, the ability to perceive spatial relations and to instinctively apply spatial transformations is likely necessary for an organism to use the information from its reflection in order to coordinate its movements and successfully touch the ``X" on its forehead.

It should be noted that these arguments assume that it is the organism's actual body parts, and not the reflected body parts, that are directly associated with the organism's self-image.  This implies that the organism is able to distinguish between the reflection of a given body part and the body part itself.  However, this does not yet resolve the issue, since the organism has to not only be able to distinguish a body part from its reflection, but be able to recognize which of the two is associated with its self-image.

There are various ways in which the organism can distinguish its actual body part from the one in the reflection as belonging to the organismal self-image.  First of all, the actual body part is distinguished from the body part in the reflection in that, in addition to being visually observed, also provides information about itself to the organismal brain via direct neural connections (a person can ``feel" their own arm, for instance).  However, this additional information can only allow the organism to distinguish between its actual body part and the reflection if the organism can correlate these internal inputs from the given body part with the spatial location of the body part.  This implies that the organismal brain is wired to recognize the various internal inputs as originating from different regions of the body, and to coordinate its movements accordingly.  Whether this ability is genetically pre-programmed or is learned is unclear.  However, it is should be noted that organisms that are not capable of self-recognition in the mirror, such as dogs, are able to scratch an itch.  This behavior is even possible with one's eyes closed.  Therefore, there is clearly some non-trivial coordination of body movements with various sensory inputs, the exact mechanism of which is likely an important area of research.

Secondly, even without internal sensory input, the organism's actual body part, say an arm, is perceived by the organismal brain from a specific vantage point, as defined by the location of the organism's various sensory organs, say the eyes, in relation to the body part.  Therefore, the actual body part and the reflected body part are perceived differently.  Because the actual body part, with its associated vantage point, is the one that is actually with the organism at all times, and is the one that is used by the organism for its various tasks, it is the actual body part that becomes directly associated with the organismal self-image.  Of course, internal sensory input strengthens the association between the actual body part and the organismal self-image, since it provides additional information that allows the organismal brain to more sharply distinguish what inputs are always present and correlated with the organism performing its various tasks.

As with visual self-recognition, voice recognition also emerges via an associative process.  The motion of the organism's mouth and the vibrations of the vocal cords produces a sound.  This sound is therefore correlated with the organism's mouth and vocal cords, or, more precisely, with the mouth and vocal cords that are associated with the organism's self-image.  Therefore, the sound correlated with the mouth and vocal cords associated with the organism's self-image itself becomes associated with, and therefore a part of, the organism's self-image.  Therefore, the organismal brain can learn to recognize the sound of the voice generated by the organism.

\subsection{The concept of ``I/Me"}

With the emergence of language, an organism capable of self-awareness is able to express its
self-awareness via the concepts of ``I" or ``Me".  Specifically, these self-referential terms become 
associated with the self-image formed inside the organism's brain, and therefore with the self-inputs that produced the self-image, and, in the case of self-recognition in the mirror, with the organism's reflection.
Therefore, as with memory, the emergence of language allows for the formation of new sets of neural pathways that are associated with, and become a part of, the organismal self-image.

To understand the emergence of ``I/Me" in self-aware organisms, we must first develop a canonical definition of these terms, one that does not rely on the concept of self-awareness.  We do this via the following example:

Suppose that an organism feels hunger, notices that another organism has food, and communicates the desire to obtain food from the other organism.  If there are other organisms around, then the organism desiring food must communicate that it is the organism that desires food, and not some other organism.  One way that this can be accomplished is if every organism has a label.  By using the appropriate label, say, ``Sam," the organism can communicate that it seeks to obtain food from the other organism, e.g. via the command ``Give Sam food".  In this case, ``Give Sam food" is associated with the organism eating the food, which is associated with the various aspects involved in eating the food connected to the organism (placing the food in the mouth, grasping it with the hands, etc.).

Therefore, because the various aspects of the organism are associated with the self-image, ``Give Sam food" is associated with the self-image as well.  Eventually, because the label ``Sam" is used in a variety of other contexts, the label ``Sam" becomes associated with the organism's self-image, and therefore with the various aspects of the organism that led to the formation of the self-image.

Now, the system of using organism-specific labels in inter-organism communication can be quite cumbersome, since it requires storing a list of the various labels in the population and the specific organisms to which they refer.  For a small population, this might not be a problem.  However, for a large, dynamic population, maintaining a long list of labels can be highly inefficient.  

As a solution to this problem, one can use canonical, temporary labels during inter-organism communication.  The organism transmitting the information is assigned the temporary label ``I" or ``Me", depending on the context (or ``We/Us" when speaking for a collective), while the organism receiving information is assigned the temporary label ``You".  Organisms not involved in the given communication
may be assigned generic labels such as ``He", ``She", etc.

There is a slight ambiguity with this definition, because, if two organisms happen to speak simultaneously, then both are transmitting information, and both are receiving information, so that the terms ``I" and ``You" can refer to either organism.  To resolve this issue, we define an inter-organismal communication to not only include the message itself, but to also include the organism transmitting the message, and the organism receiving the message.  Expressed mathematically, this means that an inter-organism communication is defined by the ordered triplet $ (\mbox{Organism}_1, \mbox{Organism}_2, \mbox{Message}) $, where $ \mbox{Organism}_1 $ is the organism transmitting the information, $ \mbox{Organism}_2 $ is receiving the information, and $ \mbox{Message} $ is the actual message transmitted.  The words ``I" and ``You" in $ \mbox{Message} $ then refer to $ \mbox{Organism}_1 $ and $ \mbox{Organism}_2 $, respectively.  Note then that this definition of an inter-organismal communication resolves the amibguity, because when $ \mbox{Organism}_1 $ speaks to $ \mbox{Organism}_2 $, the inter-organismal communication is given by, $ (\mbox{Organism}_1, \mbox{Organism}_2, \mbox{Message}) $, while when $ \mbox{Organism}_2 $ speaks to $ \mbox{Organism}_1 $, the inter-organismal communication is given by, $ (\mbox{Organism}_2, \mbox{Organism}_1, \mbox{Message}) $.  It is interesting to note that this definition allows us to formally define ``I" and ``You" for an organism speaking to itself, since such a communication is denoted by $ (\mbox{Organism}, \mbox{Organism}, \mbox{Message}) $, and so ``I" can refer to the organism in the first slot, and ``You" can refer to the organism in the second slot.

Within this system, an organism expressing a desire for food from another organism can say ``Give me food," without needing to use a specific label.  This has the advantage that the organism receiving the information understands that the food is meant to go to the organism transmitting the information, and does not require additional information that associates a label with the organism.  Then, following a similar argument to the previous one based on label-specific communication, it is possible to see that the terms ``I/Me" become associated with the organism's self-image and the various aspects of the organism that led to the formation of the self-image.

As with the phenomenon of self-recognition in the mirror, the phenomenon of ``I/Me" requires some further development beyond the basic mechanistic framework outlined here.  The main issue is that, when an organism feels hungry and decides to communicate that it desires food, it must know that the organism that is hungry is the same as the organism that is communicating.

When an organism feels hungry, it is pre-programmed to obtain and eat food.  As a result, the sensation of hunger and the act of obtaining food become associated with the various parts of the organism that are involved with eating the food (the hands, mouth, etc.).  Since these parts of the organism are a part of the organismal self-image, the organismal brain can learn, through a process of associative learning, that the process of obtaining food and satisfying an appetite is accomplished by feeding the organism associated with the self-image.  Therefore, when the hungry organism wants to communicate that it wants food from another organism, the organismal brain knows that it must use a label that corresponds to the organismal self-image.  

Now, when the hungry organism is about to communicate that it wants food from another organism, certain neural pathways are activated that, while they do not directly correspond to the organism speaking, they contain relevant information that is then subsequently sent to the neural pathways controlling the vocal cords for communication.  Since these neural pathways are associated with the organism moving its mouth and speaking, and since the organism's mouth is associated with the organism eating food and hunger, then the organism doing the speaking is associated with the hungry organism desiring food.  Therefore, the organism doing the speaking is the same as the organism desiring food, and so the organismal brain knows that providing the speaking organism with food will automatically provide the hungry organism with food.  This of course implies that the organism should use the canonical labels ``I" or ``Me".

Another set of concepts related to ``I" or ``Me" are the various concepts of possession, such as ``my" and ``mine".  An expression such as ``my hand," for example, is meant to denote the hand that is associated with the organism that is speaking.  

The concept of ``myself" is more subtle:  ``Self" is a general term that refers to a set of behaviors characterizing ``sentient" beings as distinct from non-``sentient" beings.  The term, ``Myself" is therefore meant to refer to the set of ``sentient" behaviors associated with the organism that is speaking.  This is a subtle point that will be further discussed in the following section.

While the concepts of ``I/Me" may be defined without reference to self-awareness per se, it is nevertheless an issue as to whether self-awareness is required in order for an organism to learn to properly apply these concepts.  For example, one way that an organism can learn to use the concept of ``I/Me" is to observe other organisms.  Initially, an organism might conclude that ``I/Me" is a label that refers to a given organism, until it notices that many organisms use this label, and that an organism with the label ``I/Me" in one context has the label ``You" in another context.  After extensive observation, an organism with sufficient powers of abstraction and association may conclude that ``I/Me" refers to the organism transmitting the information.  Following a similar line of reasoning to the one described in connection with obtaining food, it is then possible to see that an organism that has learned the concept of ``I/Me" will, via mediation of the associations defining the self-image, be able to use these terms appropriately.

\subsection{Awareness of self-awareness}

Another phenomenon associated with self-aware organisms is something that can be referred to as ``awareness of self-awareness".  Whether other animals experience this phenomenon is unclear.  However, humans are known to wonder why they are the way they are, that is, why they have the body and the mind that they have, and why they could not be someone else.

The question as to why a person is the way they are is not a scientific one, since, as a biochemical machine, a person's identity is defined by their physical make-up.  Such a question therefore belongs in the realm of metaphysics or theology.  Indeed, awareness of one's self-awareness has been one of the central paradoxes of human existence.  It has been a central focus in nearly every major religion and philosophy.  It is intimately connected with the mind-body problem and the concept of a metaphysical ``soul" \cite{SELFAWARE1}.  

It may therefore seem strange that a biochemical machine should develop questions that, from a scientific point of view, do not make sense and are therefore unanswerable.  However, if we view the emergence of self-awareness as a consequence of associative memory and learning in a sufficiently complex brain, then we can rationally speculate on the emergence of this phenomenon.

A self-aware organism develops an internal ``self-image'' that causes it to recognize itself in the mirror, and behave in other ways that are associated with self-awareness.  However, such a self-aware organism does not exist in isolation, but rather in a community with other organisms.  Taking humans as an example, then presumably, when one human looks at another, the second human is sufficiently similar to first human that the sensory input from the second human is associated with, and therefore stimulates, the self-image in the brain of the first human.  However, the second human and the first human are not the same, and therefore the sensory input from the second human is recognized as similar to, and yet simultaneously different from, the sensory input that led to the formation of the first human's self-image.  In an organism with a natural curiousity, the internal contradiction of having a sensory input that stimulates the internal self-image of an organism, and yet is not derived from the organism, may cause the organism to view its existing self-image as arbitrary, and to wonder why it has the self-image that it does (since in a sense, the stimulation of its self-image by another organism corresponds to an effective, temporary replacement of its existing self-image with the image of another organism).

Another manifestation of the phenomenon of awareness of self-awareness is the fact that self-aware organisms can deduce the concept of self-awareness.  Presumably, this emerges because a sufficiently intelligent organism is capable of creating complex associations that define fairly abstract concepts.  Therefore, an organism, by observing itself, perhaps other organisms, and other physical objects, may deduce the general concept of a physical object.  Essentially, the organism associates various physical objects with one another, and this association is what defines an abstract concept called ``physical object".  

However, a sufficiently intelligent organism will be able to note that not all physical objects, defined as inputs that activate the set of neural pathways that store the concept of a general object, are the same.  That is, two physical objects may stimulate a general set of ``physical object"-recognizing pathways, and two additional sets of pathways, each corresponding to the unique features of each object.  Therefore, a sufficiently intelligent organism may notice that the organism corresponding to its self-image, as well as other self-aware organisms, are physical objects, and yet they are physical objects that behave in a manner different from other objects, such as a tree, or a rock, for instance.  Therefore, a sufficiently intelligent organism may attempt to develop a concept that characterizes the set of behaviors that distinguishes between self-aware organisms and other physical objects.  If the organism first develops a general concept describing behavior whereby an organism seems to focus its attention on some other physical object, then such an organism might develop the concept of ``aware".

However, a sufficiently intelligent organism might notice that some aware organisms may not be able to exhibit certain behaviors, such as self-recognition in the mirror.  This set of behaviors that distinguishes self-aware organisms from other organisms may be attributed to a property, called ``the self", that self-aware organisms possess.  Since self-recognition in the mirror is a set of behaviors that can be observed to be a form of awareness, a sufficiently intelligent organism may attempt to refine the concept of ``awareness" to describe this particular set of behaviors.  If the organism is itself self-aware, then the organismal brain may notice that the organism associated with the internal self-image is aware of the organism associated with the internal self-image, and therefore engages in a set of behaviors defining a ``self".  In order to express this observation in language, the organism may conclude that it is ``myself-aware".  Since other organisms may be observed to exhibit behavior that is consistent with the behavior of the organism associated with the self-image, the organism may be able to conclude that other organisms are ``myself-aware".  Eventually, since the ``my" is redundant in a certain sense, the generic term that could emerge is simply ``self-awareness", though strictly speaking, one could define and distinguish between concepts such as ``myself-awareness" and ``yourself-awareness".

\section{Existential and Religious Concepts Associated with Self-Awareness}

In this section, we discuss a number of aspects of self-awareness that are apparently unique to humans.  In particular, we show how the mechanisms of associative memory and learning, combined with language, can facilitate the emergence of various existential and religious concepts associated with self-awareness.  Furthermore, we show how tautologies invariably emerge when discussing existential and religious issues associated with self-awareness, which prevent an unambiguous resolution to these various issues.

If these speculations prove to be largely correct, then it suggests that the various existential and religious questions attributed to self-awareness in humans may not be intrinsically unique to humans.  That is, other organismal lines could themselves evolve to a sufficiently high level of intelligence that they would exhibit similar behaviors.  Furthermore, it suggests that it is in principle possible to construct self-aware machines (though, due to the tautologies that emerge, this will always be an issue open for debate).

\subsection{The concepts of ``I am" and ``I exist"}

The statements ``I am" or  ``I exist" have been regarded as key signatures of self-awareness.  However, since ``I am" and ``I exist" are similar statements, we will focus our attention on the statement ``I am."

The emergence of the concept of ``I" and its association with the organismal self-image has already been discussed.  Therefore, we must next turn our attention to the concept of ``am," which is the present tense conjugation of the verb ``to be."

The verb ``to be" is used to convey information regarding the state of an object, e.g. ``The car 
{\it is} in Atlanta."  It should be noted, however, that in the present tense, the verb ``to be" is
superfluous.  That is, the statement ``The car in Atlanta" clearly conveys the state (in this case positional) information of the car.  Indeed, some languages, such as Russian and Hebrew, do not have a present tense conjugation of the verb ``to be."  Therefore, in such languages, the equivalent of the statement ``The car is in Atlanta" would exactly be "The car in Atlanta."  

When we incorporate the concept of time, however, then the verb ``to be" becomes necessary, in order to assign a temporal label to the state information.  One label can be used to indicate past
state information, while another label can be used to indicate future positional information. 
Therefore, the statement, ``The car {\it was} in Atlanta," indicates that a given car was located in Atlanta
at some undefined point in the past, while the statement ``The car {\it will be} in Atlanta," indicates
that a given car will be located in Atlanta at some undefined point in the future (in the English language, these statements do not necessarily preclude the car being in Atlanta in the present.  Different languages have different subtle variations in tenses to handle various cases).

Note that the words ``was" and ``will be" are not superfluous in their respective sentences.  They convey additional temporal information that ``The car in Atlanta" does not.  Indeed, even languages such as 
Russian and Hebrew, which do not have a present tense conjugation of the verb ``to be," exactly use this verb to convey both past and future state information.  Therefore, in Russian and Hebrew, the equivalent to the statement, ``The car was in Atlanta" is exactly ``The car was in Atlanta."

The present tense conjugation of the verb ``to be" could then emerge to fill the gap between the past and the future tenses of the verb.  That is, other verbs, such as ``to throw, to eat, to sleep, to walk, etc." require present tense conjugations.

From this mechanistic perspective, the statement ``I am'' is seen to be a tautology, that is, true by
definition.  The term ``I'' refers to the organism that is speaking the ``I", and hence the organism's self-image, and ``am'' has no intrinsic meaning except in the context of referring to the state of an object.  

Similarly, the statement ``I exist'' is also a tautology.  To say that an object exists is to state that it is a construct of the physical universe.  Since ``I'' refers to a self-image that was produced by various inputs associated with a given organism, ``I'' refers to a construct of the physical universe, and therefore,
``I exist'' is true by definition.  

\subsection{The concepts of ``mind'', ``soul'', and ``self"}

Another set of related concepts that emerge in sufficiently intelligent organisms are the concepts of ``mind", ``soul", and ``self".  As hinted at previously, the emergence of these concepts require an organism to associate contradictory constructs with organisms that it observes, including the organism that is associated with the self-image (i.e., the organism itself).  

Specifically, a sufficiently intelligent brain may develop the general concept of a physical object, with which it associates other organisms, as well as the organism associated with the self-image.  At the same time, the sufficiently intelligent brain may notice that different physical objects exhibit different behaviors, so that certain physical objects, such as humans, apes, and elephants, say, are recognized as distinct from other objects, such as rocks and trees.  The result is that the set of behaviors distinguishing one set of physical objects from another may provide the basis for defining a new concept.

In the context of self-aware organisms, the set of behaviors distinguishing such organisms from non-self-aware organisms may collectively be used to define a concept, which may be termed the ``mind", ``soul", or ``self".  Because such concepts describe a set of behaviors not shared by all physical objects, these related concepts become entities that are distinct from physical objects.  Therefore, the notion of ``mind", ``soul", and ``self" may become regarded as entities that are associated with self-aware organisms, but are simultaneously distinct from the physical make-up of the organisms themselves.

Various religions have then developed differing speculations as to the relation between the ``soul" and the physical organism.  Some religious traditions argue that the ``soul" is intrinsically tied to the physical organism, and the destruction of the physical organism leads to the destruction of the ``soul".  Other traditions argue that when an organism with a ``soul" dies, the physical make-up of the organism is destroyed, but the ``soul" is preserved \cite{RELIGION}.  To understand the basis for such a view, we note that if an organism dies by being killed, this generally involves some other external physical object affecting the organism in a way that causes it to cease functioning.  However, since the external physical object can affect other objects that do not have a ``soul" (a bullet can put holes in both living tissue and in rocks, for example), one conclusion that can be drawn is that these external physical objects can only afffect the physical aspect of an organism, but that the set of behaviors defining the ``soul" are unaffected.

Such a view is likely reinforced by the innate fear of death.  The fear of death can trigger an internal stress response inside the organismal brain, which drives the formation of neural pathways that can suppress this stress response.  Since a dead organism is recognized as being a physical object that appears similar to the organism when it was alive, the fear of death, in a sufficiently intelligent organism, may lead to a stress response that causes the organism to attempt to preserve the set of behaviors characterizing the ``soul".  The conclusion that death only affects the physical aspect of the organism, but the non-physical ``soul" is preserved, defines a set of neural pathways that suppress the internal stress response in the organismal brain.  Because this conclusion is an internally consistent one, it does not lead to any contradictions in the organismal brain that can lead to an ``un-wiring" of the neural pathways.

Note that the existence of a mind, a metaphysical soul, and the self is true by definition.  As with the concepts of ``I am"/``I exist", these concepts are tautologies by the nature in which they were constructed.  Furthermore, the various interpretations associated with these concepts are also tautologies.  The reason is that all of the concepts developed here, as well as the various interpretations, were developed by developing distinct sets of concepts to describe a certain class of organisms (the recognized self-aware ones).  Since these concepts are generated by such organisms, they are intrinsically associated with one another, and therefore it is not a scientific question to speculate as to whether these concepts can exist separately from one another.  

\subsection{Solipsism}

With the concepts of ``mind", ``soul", and ``self" in hand, we can now proceed to speculate about several other behaviors that are associated with sufficiently intelligent self-aware organisms.  These behaviors are connected to the idea that reality is, in some sense, an illusion.

One such behavior is known as {\it solipsism}, and essentially means ``myself alone".  In the solipsistic perspective, a given organism is the only self-aware being in the universe.  Furthermore, as far as that organism is concerned, the universe only exists as long as that organism is there to observe it.

To understand how an organismal brain may develop such a conclusion, we need to first analyze various kinds of associations that the brain forms.  To begin, when the brain receives sensory information from the external universe, it receives information about the organism in which it sits, as well as information from the environment (strictly speaking, from the point of view of the organismal brain, all signals may be viewed as ``environmental'' signals).  Furthermore, if the organism has a long-term memory that can be stimulated, then the organismal brain can have simultaneous activation of various neural pathways that correspond to all the various aspects of universe that the organism has encountered in its lifetime.  Such activated neural pathways can become associated with one another, which can then be used to define a neural pathway that is a representation of the general concept of reality.  However, as a result of the associations formed, the organismal self-image is a part of the associations defining the physical reality.  Therefore, the organismal brain associates its self-image with the external world.  

The existence of sleep and death, however, complicates the association between the organismal self-image and the rest of the physical universe, in such a way that it results in a number of equally valid associations, in the sense that none of them may be proven true or false.

To understand this, consider the following thought experiment:  A human observes the motion of the sun through the sky, and eventually forms in its long-term memory a set of associations that corresponds to this motion.  Now, suppose the organism observes the sun at some position in the sky, and then decides to take a nap.  The organism may form a memory of the sun at some given position in the sky, as well as a memory of the organism lying down, and closing its eyes.  Once the organism falls asleep, its next set of memories will be of it opening its eyes, getting up and feeling groggy, and noticing that the sun is at some new position in the sky.  

Therefore, the organism forms a memory of the sun in one position in the sky, and then in another position.  Although the organism does not have a memory of the continuous motion of the sun in this particular instance, previous experience may cause the organism to ``fill the gap" and to infer that the sun was moving through the sky, and the lack of this memory is associated with the various features of the organism going to sleep that the organism remembers (closing the eyes, lying down, feeling groggy, etc.).  However, although the organism may infer that the sun should have been in a continuum of positions as it moved through the sky, the organism has no memory of this series of events, nor of itself in the instances when these events occurred.  Since the organism associates its self-image with the external reality, there is an apparent contradiction in a series of memories of the physical reality that do not involve the organism.  Therefore, since the organism does not have a memory of itself during the intervening positions of the sun, the organism may conclude that there was no external reality either during the time it was asleep.

As a result, the principle of ``correlation implies causation", combined with the association between the organismal self-image and the external reality, may cause the organism to conclude that the physical world only exists when the organism is there to observe it.  Equivalently, the organism may conclude that it creates the physical world, including all the other organisms in it. 

A similar argument can be achieved by considering death, although it is somewhat complicated by the fact that an organism cannot observe its own death.  In this case, the organism must observe that other organisms die, and infer that it too will die.  By noticing that other organisms that are dead cannot observe the universe, the organism may infer that it too cannot observe the universe either when dead, which may also lead to a solipsistic view.

\subsection{The idea that thought creates the physical world}

Another notion related to solipsism is the idea that thought itself creates the physical world.  Such a view also has a number of forms, but emerges as an attempt to explain the apparent contradiction in the solipsistic view that, on the one hand, a given organism creates the physical world, and yet at the same time is a physical being as well.  

This apparent contradiction may be resolved by viewing the organism itself as a part of the physical world, that exists only when the organism is there to observe it.  But this is a problem, for if the organism is not there in the first place, then it cannot observe and create itself.

However, by the construction of such concepts as the ``mind", ``soul", and ``self", the organism can make a separation between the so-called physical aspects of the organism, and the set of behaviors that define a new abstract, metaphysical concept.  The idea in this case is that the physical aspects of the organism are indeed a part of the external reality, which only exists if the metaphysical aspects are there to observe it.  However, because the metaphysical aspects are not associated with the physical reality, they are not presumed to disappear when there is no input from the sensory world.  As a result, a logical conclusion that follows is to accept the notion of ``mind" as a primitive, that exists independently of any external reality.  Because the organism opening its eyes leads to observation of the world, and because the opening of the eyes is associated with the set of behaviors defining a metaphysical ``mind", the logical conclusion that follows from this is that the universe is created by the ``mind" observing the world.

However, even here we have an apparent contradiction, because observing the world involves the use of organs that are associated with the physical universe.  This is resolved if the organism can be aware of itself thinking as opposed to actually observing the world.  Such a distinction should be possible because, during thought, certain constructs are formed via the stimulation of certain neural pathways.  However, during thought, these constructs are formed independently of any external stimuli, so that certain neural pathways that are responsible for interaction with the environment (muscles, sensory organs, etc.), are inactivated.  The result is that the process of thought leads to a somewhat different set of associations than the process of actual observation and interaction with the environment.  If the organismal brain can translate this distinct set of neural constructs into a word, i.e. ``thought", then the organismal brain may note that ``thought" involves the manipulation of representations of physical constructs without interaction with the environment itself.  Since ``thought" does not involve explicit interaction with the environment, it is not directly associated with the physical reality.  Furthermore, since ``thought" is associated with the ``mind", being a member of the set of behaviors not characteristic of other physical objects, the organism may conclude that there is an abstract ``mind" that is capable of ``thought" that exists independently of any physical reality.  However, since ``thought" involves representations of physical reality, then the organism may conclude that the ``mind" creates the physical world via ``thought".  This is a more abstract form of solipsism, but a form of solipsism nonetheless.  As with previous concepts discussed in this section, it is a view of physical reality that allows the organism to ``cheat'' death, and so it is a view that is likely reinforced by the internal reward-punishment-based neural pathway formation processes at work in the brain.

\subsection{Reincarnation}

Reincarnation is another common notion that humans have developed.  The origin of this idea comes from the association of the organismal self-image with the external reality.  The solipsistic view argues that the organism creates the reality, and that reality itself is actually an illusion.  However, an organism may be uncomfortable with this conclusion, since this implies that other other, similar organisms are an illusion that the organism creates.  There may therefore appear to be something arbitrary about the reality having been created from the vantage point of an organism with a particular self-image.

One solution is to reject the notion that reality is an illusion.  The reality in fact exists independently of any particular organism.  However, since the organism views the reality, as well as its self-image, from a certain vantage point, and since this vantage point is associated with the reality, the organism may conclude that the vantage point from which it views the reality is a generic one by which the reality is viewed.  If the reality does not disappear when the organism dies, then all that changes is the vantage point, so that the reality will be viewed from the perspective of another organism.  This is essentially the content of reincarnation.

One version of reincarnation holds that there is only one ``mind" or ``soul", and that this entity is distributed over many different organisms, in a kind of quantum superposition.  This view emerges because reality is viewed from the perspective of a single organism, and hence from a single ``mind".
The other organisms in this reality, since they are not the perspective from which the reality is being viewed, are then simply physical objects that do not have a ``mind" or ``soul".  However, it may then seem somewhat arbitrary that only the given organism has a ``mind", and so one solution to this apparent contradiction is that there is a single ``mind" in the universe, and so the reality can only be observed from the perspective of one organism at a time.  However, every organism has an equal probability of being the vantage point of observation.  This view is consistent with the many-worlds interpretation of quantum mechanics:  For every organism, there is a universe where the reality is viewed from its perspective, and in that perspective, all other organisms are simply biochemical machines.

\subsection{The concept of divinity}
 
The final religious concept we shall discuss is the concept of divinity.  The concept of divinity, especially the monotheistic notion of a transcendent, all-powerful deity, emerges by entirely disassociating the concept of ``mind" from the physical aspect of the organism.  The idea is that, because the ``mind" is a nonphysical aspect of the organism, then it must exist independently of the organism.  However, unlike the previous discussion, the ``mind" does not need a physical instantiation in order to express itself, but rather can exist and function fully independently of any physical representation.

Furthermore, by noting that it is the ``mind" aspects of an organism that allows the organism to manipulate other physical objects and create tools, the organism might reason that the entire physical reality may have been created by a ``mind" entity that has no physical form, since it itself creates the reality.  In this view, since humans are physical beings that engage in a set of behaviors that is indicative of possessing a ``mind", certain religious views hold that humans were created in an image, or likeness, of this transcendent, omnipotent ``mind".  In a sense, humans then contain a kind of projection of this ``mind," that is of course far less powerful because it is bound within a physical body.

As a final note, the concept of divinity as a single, transcendent, omnipotent ``mind" that creates the universe, combined with the idea that there is one ``mind" superposed over many organisms, and the idea that thought creates the physical universe, may lead to the identification of the single, superposed ``mind" with the transcendent ``mind" that creates the universe, and therefore to the conclusion that every human is an instantiation of this transcendent ``mind".  Indeed, in sufism, or Islamic mysticism, there is a concept that each human is in a sense God, that is, also divine \cite{RELIGION}.  This idea emerges precisely from this set of identifications.  Presumably, other identifications could lead to other concepts.
 
\subsection{Tautologies and the nature of reality}

The list of the various religious and existential concepts described above is not meant to be exhaustive, but rather representative of various streams of modern religious thought.  Note that all of these religious concepts arise from the simultaneous association and disassociation of an organismal self-image with the external reality.  This leads to a variety of contradictory identifications, and the attempt to resolve these contradictions leads to a variety of additional concepts or constructs.

What must be noted is that all such concepts are tautologies, that is, true by definition.  Because all of these concepts emerged from the way in which the organism's sensory organs observe the organism and the external reality, there is no way to either prove or disprove the validity of these constructs.  
In the language of set and logic theory, one would say that these various existential and religious questions are formally {\it undecidable} \cite{SETTHEORY1, SETTHEORY2, HALTING}.  Indeed, we argue that the various existential and religious questions associated with self-awareness have remained unresolved precisely because they are ultimately tautologies, and therefore unresolvable.

The fact that these various constructions are tautologies means that any explanation for the emergence of self-awareness must deal with the possibility that the external reality is in fact a creation of the self-aware organism.  This of course implies that an organism can never fully establish that the explanation given for other organisms' self-awareness is an explanation that applies to the organism itself.  That is, self-awareness and consciousness, and therefore the notion of a ``mind" or ``soul", may in one tautologically valid view, need to be taken as primitives that have no explanation in terms of other phenomena.  This implies that we may then ask what is the utility of attempting to provide a mechanistic explanation for the emergence of self-awareness.  To formulate a proper answer to this question, we must first discuss the general manner by which physical laws are deduced.

Physical laws are essentially compactified representations of an entire collection of physical phenomena.  To illustrate, Newton's Second Law of Motion, $ F = ma $ may be viewed as a compact way of describing a continuous collection of phenomena, where each phenomenon is an instantiation of a force acting on a mass that produces an acceleration.  Therefore, strictly speaking, it does not make sense to ask whether physical laws are true or not, in the sense that their usefulness comes not from any notion of a higher truth, but rather whether or not they provide a convenient framework for describing physical phenomena, in a manner that facilitates the prediction of the behavior of systems, and the construction of systems based on these behaviors.

Similarly, it does not make sense to ask whether the speculations in this paper are true or not, for, as we have shown, depending on how the various abstract concepts are defined from a given set of behaviors, and how a given question is framed, the speculations may be shown to be either true or false.  Rather, this paper has identified associative learning as the underlying mechanism for the emergence of self-awareness in big-brained organisms.  The utility of this framework will not be its ability to resolve tautological issues such as the nature of the self, the soul, or the mind, but rather whether this framework
will facilitate the development of predictive models for self-awareness, including models for the evolutionary biology of the phenomenon, and the development of machines that can exhibit behaviors associated with self-awareness.

Ultimately, then, from a scientific perspective, it does not matter whether there is an objective reality that exists independently of the organism observing it, or whether the reality is created by the organism itself.  In the former view, the objective reality is governed by physical laws that produced self-aware organisms capable of manipulating the reality and deducing the physical laws by which the reality is governed.  In the latter view, the solipsistic reality (the one created by the organism) is also governed by physical laws that produced self-aware organisms capable of manipulating the reality and deducing the physical laws by which the illusory, solipsistic reality is governed.  Since these two views are both tautologically valid, neither can be falsified by any kind of experiment, and debates about the correctness of one view over the other is relegated to the realm of philosophy (it should be noted that we are referring to the philosophical, existential notion of an ``objective reality", so that this discussion does not deal with the so-called ``measurement problem" in quantum theory).

\section{Manipulating Self-Awareness}

In this section, we speculate on several possible ways to manipulate self-awareness.  We emphasize that these speculations should be viewed purely as thought experiments, and not as suggestions for actual experiments on big-brained organisms. 

\subsection{Sensory inhibition and mathematical modeling}

One approach to test whether or not self-awareness is a learned behavior that emerges from the formation of a self-image is to see whether the emergence of self-awareness can be controlled by regulating the sensory inputs responsible for forming the self-image in the first place.  Essentially, by inhibiting the input of sensory information during the early stages of brain development, when the neural pathways corresponding to the initial nucleus of the self-image are being formed, it may be possible to prevent the formation of the self-image.  However, for a sufficiently complex organism, this may be extremely difficult to accomplish in practice, because in addition to the five senses, the number of internal sensory inputs to the organismal brain may be so large that only a drastic reduction in sensory information could severely inhibit the formation of a self-image.

A related set of studies involves monitoring the development of neural pathways corresponding to a self-image in infant brains.  Such studies could also be complemented by mathematical modeling that, based on underlying mechanisms for neural pathway formation, could attempt to predict the various times at which various general features of the self-image emerge.  In this vein, one interesting question is whether the formation of a self-image involves a kind of percolation phase transition.  That is, is self-awareness a gradually emergent phenomenon, or is it characterized by the formation of a critical nucleus of neural pathways defining a core self-image?  This core self-image would provide a threshold level of associations between various organismal features, drastically facilitating the processing and integration of further information related to the organism.

If self-awareness does involve a percolation-like transition of neural networks in the brain, then this should manifest itself in terms of organismal behavior.  For example, humans are first capable of recognizing themselves in the mirror at around 15 months of age \cite{SELFAWARE3}.  Presumably, at birth, the human baby's brain is essentially a ``blank" that requires about 15 months to develop the critical nucleus of neural pathways necessary for self-recognition.  If the phase transition picture is correct, then the onset of self-recognition in the mirror will be characterized by a binary switch behavior and will occur suddenly.

\subsection{Brain transplants}

Another approach is to study what happens when the brain of one organism is transplanted into another.  If the brain is transplanted at a sufficiently young age, before the self-image has had a chance to form, then the brain may be regarded as a ``blank" that has not formed neural pathways that contain the imprint of the organism in which it sits.  Such a brain may therefore be transplanted into a new organism with little difficulty (not including the actual surgical operations, of course, which are a major difficulty), and the brain can form the self-image based on the new organism in which it sits.

For an organismal brain that has already formed a self-image, we conjecture that the result of the brain transplant depends on the complexity of the organism.  The issue here is that the brain has already formed pathways corresponding to the neural imprint of a certain organism.  These neural pathways corresponding to the self-image therefore respond to, and essentially ``expect" these inputs, in the sense that the absence of these inputs can cause the neurons forming the pathways to begin a re-wiring process in the search for inputs that they can respond to.  This in turn can cause the organismal brain to trigger a stress response.

However, depending on the organism, the self-image may be more robust to brain transplants in some organisms than in others.  Presumably, a sufficiently complex organism may be capable of forming large sets of associations that correspond to various abstract concepts.  The result is that an organism such as a human may store a more abstract form of the self-image than a simpler organism, such as an elephant.

Therefore, a human brain may be transplanted with relatively little difficulty.  While the brain might initially experience a transient stress response, due to the slightly different sensory information it is receiving, the self-image may be stored in a sufficiently abstract form that the brain may be able to adjust its internal wiring to recognize the new organism in which it sits.  We should note that, by a more abstract form, we mean that the self-image stores the general vantage point from which it views the organism and the external reality, and recognizes general features of the organism (such as an arm, leg, etc.), without representing these features in great detail.

A simpler organism, such as an elephant, may store the self-image in a more literal manner.  The result is that when the brain of such an organism is transplanted, it may be unable to respond to the inputs coming from the new organism.  The organism in which the transplanted brain sits may then seem extremely stressed and confused, and may take a long time to reach normal behavior, if ever.

The additional advantage for a more complex organism, especially one that is capable of communicating via language, is that it can know ahead of time that its brain will be transplanted.  The brain can therefore prepare itself ahead of time, by forming temporary neural pathways that will allow for a smoother interface between itself and the new organism in which it will sit.  The advantage of language is that these neural pathways could possibly be routed through the language centers of the brain, which could provide a more abstract ``emulation" of the self-image that could facilitate integration into the new body.

Of course, a simpler organism may, in any case, store a less detailed self-image than the more complex organism, suggesting that a simpler organism may actually have an easier time adjusting to a brain transplant than a more complex organism.  Indeed, in organisms that are so simple that their brains are not capable of forming a self-image at all, brain transplants should have little effect on organismal behavior.  Clearly, there are competing effects involved, which likely leads to some intermediate brain complexity at which brain transplantation becomes most difficult.

\subsection{Brain-body separation}

Another manipulation of self-awareness that we believe will be possible is the physical analogue of mind-body separation, what is more appropriately called {\it brain-body} separation.  The idea behind this concept is that it should in principle be possible to remove the brain from an organism's body in such a way that the behavior, and hence self-awareness, of the organism is not affected at all.

Suppose the brain of a self-aware organism is extracted from the organism, in such a way that all of the connections between the brain and the rest of the body are maintained, and suppose the brain is kept alive and functional.  Because the brain is still processing the external and internal sensory information in the same manner as before, the organism will still continue to function as normal, and the neural pathways defining the organismal self-image will be unaffected.  Essentially, the organism continues to view the world from the same vantage point, which might seem strange, since awareness is located in the brain, and yet the vantage point does not change to one corresponding to the brain viewing the organism from outside its body.

Now, suppose that the various inputs to the organismal brain could be accomplished via some kind of wireless connection, so that the organismal brain could be completely physically detached from the organism itself.  In this situation, the organismal brain would still be processing the sensory information in the same manner, and so again, the self-image would be unaffected and the organism would appear to function normally (though it might be able to view its own brain).

What would be achieved in such a situation is literally brain-body separation.  The organism could have its brain kept alive and stored in some location, and kept in contact with the organismal body via some kind of radio frequency connection.  Because the organismal brain is receiving sensory information from the organism and the external world from the same set of sensory organs, with the relations among these inputs preserved, the organismal self-image is unaffected by this brain-body separation, and the organism will still feel as if it is perceiving the world and itself from the same perspective. 

In this situation, if the organismal body is destroyed, the brain-body connection will be destroyed as well, and so the organism will be effectively dead.  However, if a brain-body connection is established with another organism without a brain-body connection, then the new organism will have the memories corresponding to the old organism.  In a sense, then, what would have been accomplished is a physical realization of the phenomenon of reincarnation.  

\subsection{Multiple bodies in a single mind}

What if two or more sets of organismal inputs were fed into a single brain?  It is very possible that the brain would develop a kind of ``composite" self-image corresponding to the two set of bodies.  Correlated behaviors between the two bodies could emerge, even if the bodies are spatially separated by vast distances.  The light-speed barrier to information transmission would, however, force a temporal separation to these correlated behaviors.  However, we would have a mind that is superposed over multiple bodies.  This mind could experience spatial delocalization, in the sense that it could focus its attention on some aspect of one organism, say looking at one arm, and then switching its attention to focus on an aspect of the other organism.  The mind would essentially feel like it was in several places at once.

\subsection{Will it be possible to construct self-aware machines?}

The issue of whether it will be possible to construct truly self-aware machines is composed of two questions:  The first question asks whether or not it will be possible to construct machines that exhibit the behaviors of other self-aware organisms, such as humans.  This question is one within the realm of scientific investigation, as it deals with physical constructs that may be manipulated.  The second question asks whether such machines would be truly self-aware, i.e., would they have a ``soul", or whether they would simply be imitations of truly conscious life.  As discussed in the subsection dealing with tautologies, the answer to this question depends on how one defines the various metaphysical constructs of ``mind" and ``soul", and is therefore not one that can be ultimately resolved.

There are two possible approaches in designing machines capable of exhibiting self-aware behavior.  One approach involves enumerating the behaviors associated with self-awareness, and explicitly programming this behavior into the machine.  This approach is probably extremely difficult, because the list of behaviors is likely so large that a direct enumeration of them would lead to prohibitive memory and processor costs.  

Another approach is to have the machine learn the behaviors based on a smaller set of rules acting on external environmental inputs.  This approach is more compact, and also has the advantage that it allows the machine to acquire new behaviors in response to varying environmental inputs.  Furthermore, because this approach is likely more closely related to the manner in which humans respond to the external environment, it will likely lead to machines that exhibit more human-like behavior.

Therefore, one can argue that it will indeed be possible to design self-aware machines.  Indeed, machines that are capable of some form of self-recognition have recently been designed \cite{ROBOT1, ROBOT2}.  Whether this implies that the machine has the beginnings of consciousness, a ``mind", or a ``soul", or is merely imitating these phenomena, is ultimately an unresolvable question.  Indeed, one could argue that something as simple as a self-replicating molecule is capable of self-recognition, since it needs to catalyze its own replication.

\section{Further Comments}

In this section, we consider a number of additional topics that are not directly related to the main points of this paper.  Nevertheless, because these topics deal with issues that emerge naturally in a discussion of self-awareness, they are worth discussing.

\subsection{The role of mirror neurons}

Mirror neurons are a recently discovered class of neurons that are currently subjects of intense investigation \cite{MIRRORNEURON1, MIRRORNEURON2}, because they are believed to play a key role in organismal behaviors such as imitation and the emergence of empathy \cite{MIRRORNEURON1, MIRRORNEURON2}.  The mirror neurons are a set of neurons that fire in response to a specific type of behavior, whether that behavior occurs in the organism itself or in another organism.  Mirror neurons are named this way because their firing causes the organism to imagine itself engaging in the behavior of another organism, and so they ``reflect" a given behavior from one organism onto another.

It can be seen that mirror neurons may also play a key role in the emergence of self-awareness, since they recognize various features of an organismal self-image.  However, what is not immediately clear is whether mirror neurons are a special class of neurons, or whether they belong to a more general class of neurons that can form associative pathways, and simply happen to respond to features associated with a given organism.  If the latter is the case, then self-awareness is not an emergent property of a special class of ``mirror neurons," rather it is the result of a set of associations that are implemented by a general class of neurons, which are then termed ``mirror neurons" because they implement those associations. 

To illustrate this point, consider a set of neurons that, either individually or as part of a pathway, recognize eating behavior in an organism.  These neurons will then fire when the organism in which these neurons sit eats, but then will also fire when the same organism sees another eating.  Therefore, one might argue that these neurons were pre-set to recognize organismal behavior.  However, physiologically, they may be no different from neurons that are involved in other conditional responses, for example in the case of Pavlov's famous series of experiments with dogs.

\subsection{Parental and other forms of awareness}

The argument that self-awareness is fundamentally a learned behavior faces a potential difficulty, in that before many self-aware organisms become self-aware, they exhibit a form of awareness called ``parental awareness," where the organism recognizes its parent, but does not yet recognize itself \cite{PARENTAWARE}.  Since recognizing another organism might seem like a more difficult task than recognizing oneself, it might seem that if self-awareness were a learned behavior, then it should emerge before parental awareness, and not after.

To resolve this issue, we begin by first noting that some aspects of self-awareness must be hard-wired.  That is, the organismal brain must be pre-wired to recognize certain sensory connections as coming from certain parts of the body, and to coordinate its various motions with these sensory inputs.  Otherwise, the organism will not be able to form a self-image.  Furthermore, the organism must be pre-wired to engage in certain sets of behaviors in response to certain inputs, so that these two can become associated with one another to form a self-image.  As an illustration, an organism that is not pre-wired to seek and eat food when hungry will not try to learn verbal commands for obtaining food from another organism, and will therefore not learn the concepts of ``I/Me".

Therefore, at some level, the organism must have a certain amount of ``hard-wired" genetic programming that leads to the construction of a brain that, given a certain set of inputs, is capable of processing those inputs in a specific way and leads to certain forms of learned behavior.  As a side note, this implies that while some manipulations of self-awareness are possible, others will require a re-wiring of certain aspects of neural circuitry to make the brain ``compatible" with the new inputs the brain receives as a result of the given manipulation.

However, because the genetic programming of an organism itself evolved through a long process of replicative selection, it may be argued that the genetic programming itself was ``learned".  In this view, if we do not restrict the definition of awareness to include behaviors that are defined at the organismal level, but may also include ``behaviors" that emerge at the cellular and even biochemical levels, then one may view awareness as a ``learned" behavior that emerges over long evolutionary time scales.

In general, there is no canonical definition for these terms.  A behavior is only learned with respect to some pre-set rules that are hard-wired.  Those hard-wired rules may have themselves been ``learned" with respect to some other set of rules that are hard-wired on a lower level.  In principle, the chain can continue indefinitely, and so in the final analysis, we have to start with something:  Either we postulate a set of hard-wired rules that, when coupled to external interactions with the environment, lead to a set of learned behaviors, or we take as our primitive an ability to learn that, when coupled to the environment, leads to the emergence of rules that then may become hard-wired.  The viewpoint to take largely depends on considerations of convenience for the given problem at hand.  For example, when building a machine to deal with various environmental inputs (say, a stress response), the advantage of having responses hard-wired is that the machine can respond more quickly to a given set of inputs, and therefore has a higher probability of behaving effectively (one would say that such a machine has good ``instincts").  On the other hand, such a machine may not be able to effectively deal with new inputs.  By contrast, a machine that has to learn its responses may be at an initial disadvantage, since it does not have a pre-wired set of responses to a given situation.  At the same time, because the machine can learn its responses, it can optimize its behavior for a given set of inputs.

In this view, we may understand the emergence of parental awareness as preceding self-awareness in a number of ways:  One possibility is that the organism is wired with certain basic survival instincts.  That is, the organism, even at birth, feels hunger and desires food.  The problem for a human baby, say, is that its other systems have not ``come online", so that it is unable to perform the activities that allow it to acquire food.  To compensate for this, the infant human is programmed with certain basic responses, such as crying, that are activated when the organism is hungry.  Because the adult parents are programmed to respond to these cries by feeding the baby, the baby human's brain learns to associate the parent with various aspects of its own survival (obtaining food, etc.).  As with the organismal self-image, the result is the formation of a set of associations between the parent and the various basic pre-programmed survival instincts in the baby's brain, that leads to a kind of ``parental" self-image, leading to behavior that may be termed ``parental awareness".  However, because the infant brain is not as developed as the adult brain, ``parental awareness" is likely represented in a much simpler manner than self-awareness.

Eventually, as the baby grows and the brain begins to take greater and more sophisticated control over the baby organism's body, the brain is able to process the external and internal sensory information in a much more sophisticated manner than before.  Furthermore, as the organism learns to function more and more independently, the organismal brain begins to adopt behaviors and survival strategies that rely on the organism itself, and not on the parent.  The various pre-programmed instincts now lead to organismal behaviors that lead to associations that create a self-image, and the result is the emergence of self-awareness.

\subsection{The problem of qualia}

One of the central issues that arise in attempts to understand the phenomena of self-awareness and consciousness is the problem of qualia.  Specifically, self-aware organisms are seen to perceive the world using certain primitives, called {\it qualia}, as the basic building blocks for constructing reality (one of the most commonly cited examples of qualia is color, and probably the most commonly cited color is red, given the color's comparatively powerful affect on the brain) \cite{QUALIA}.

The existence of qualia raises the question as to why the world is perceived with a given set of qualia, and not others.  That is, why is electromagnetic radiation perceived in a manner that is called ``color'', while air vibrations are perceived in a manner that is called ``sound''?  Furthermore, how is it possible to establish that the manner by which reality perceived by one organism is the same as the reality perceived by another organism?

Because qualia are primitives as perceived by a given organism, they cannot be defined with respect to other qualia percevied by the same organism or by similar qualia perceived by different organisms.  As discussed previously, since science can only analyze interactions between objects, and cannot tell us what an object ultimately ``is" (a question that therefore does not make sense, scientifically), the above questions, as with the existential questions discussed earlier, are ultimately unresolvable in a scientific sense.

While we cannot resolve the problem of qualia in a scientific manner, we can nevertheless rationally speculate as to how a self-aware organism may formulate such a problem in the first place.  That is, while we cannot define what color ``is", and therefore we cannot answer whether or not two people perceive red in the same way, we can attempt to resolve how such a question would arise in the first place.

When an organism observes a color, say red, certain neural pathways that recognize the color are stimulated.  An organism with a language ability may then formulate a word (``red") to describe the color.  More precisely, the word ``red" is instantiated in the organismal brain by certain neural pathways involved in language, which are associated with the neural pathways involved in vision and color recognition.  However, because the word ``red" is itself constructed from letters, the neural pathways instantiating the word ``red" are associated with neural pathways that instantiate letters.  Since letters are not associated with the neural pathways instantiating the color red, the result is that the organismal brain forms an internal contradiction, whereby the color red is recognized as being associated with, and yet different from, the word ``red".  Since other qualia are recognized as different from one another, and yet are associated with one another via their association with letters, the organism may ask why it perceives a given qualia the way it does, or more precisely, why the qualia are not all perceived in the same manner (it should be noted that we could reach a similar set of conclusions if, instead of using language to represent a qualia, we used another set of constructs, such as those found in mathematics).

The related question of whether two people perceive color in the same way arises as follows:  When a given self-aware organism perceives the color red, the neural pathways being stimulated may be associated with other neural pathways involved in the organismal self-image, simply due to the fact that, while an organism perceives its reality, it also receives information about itself.  Therefore, the color red that the organism perceives does not exist in isolation, but rather is a construct that is associated with, and therefore intrinsically tied to the organismal self-image.  

However, a self-aware organism may associate its self-image with the image of a second organism, and in this way the neural pathways involved in processing the color red may become associated with the second organism.  This leads to two related, yet distinct representations of the color red in the first organism's brain.  The first representation is one that associates red with the organism itself, while the second representation is one that associates red with the second organism.  These two constructs are similar, and yet distinct, an internal contradiction that can cause the organism to attempt to resolve whether or not they are indeed the same or not.  A natural question that may then arise in the first organism's brain is whether the ``red" that is associated with itself is the same as the ``red" associated with the second organism.  

As with many of the constructs discussed in this paper, this is a formally undecidable question, and can be regarded as tautologically true or false depending on how one chooses to define the concept of color.  If color is defined to be intrinsically connected to the organism that is perceiving it, then, strictly speaking, different organisms do not perceive a given color in the same way.  However, if color is taken to be a concept that is distinct from the organism perceiving it, then, if two organisms can communicate to one another agreement that a given color is being observed, then by definition the two organisms perceive color in the same way (since in this case it may be argued that it is merely the representation of the color that is being translated from one form to another).   

\subsection{How is it possible to deduce physical laws?}

If it is possible to provide an explanation for the emergence of self-awareness, then it is possible for ``mind" to understand ``mind".  Like self-awareness itself, this is a phenomenon that has also been a subject of extensive philosophical debate.

To place this issue in proper context, we should note that it belongs to a more general question:   How is it that the universe, through the process of Darwinian evolution, has produced biochemical machines that are capable of deducing the physical laws by which the universe was created?

In proceeding to resolve this issue, we must first caution that, because of the problem of tautologies that emerge in such discussions, e.g. the universe is simply created by the organism itself, a resolution to this issue is ultimately impossible.  Therefore, our goal will not be to provide an unambiguous resolution to this issue, but rather to place it in the context of a useful framework.

When humans deduce physical laws, they are observing a collection of physical phenomena that they then compactify into a convenient representation.  Whatever form this representation takes (mathematics, language, a computer program), the important point to note is that the physical law itself is not really discovered, but rather is translated from one representation into another.  Therefore, the deduction of physical laws may be likened to a replication process, where a given collection of information is replicated from one representation into another.  The human organism then becomes analogous to a catalyst that catalyzes this replication process.  The problem of discovering how various aspects of reality function is reduced to a question of whether replication of various aspects of reality is possible.  That self-replication is possible is a given, since there are self-replicating systems at various levels of complexity (molecules, cells, organisms).  Furthermore, there are enzymes capable of catalyzing the translation of information from one representation into another, an important biological example being the translation of genetic information from the nucleic acid to the amino acid representation.  

Although the deduction of physical laws essentially amounts to the translation of information from one representation into another, certain representations are preferred over others.  This may be due to the fact that certain representations are more convenient, in the sense that they store the information in a form that is much easier to manipulate.  For example, since human beings are capable of self-replication, one can argue that we ``know" how to make ourselves, and therefore human beings already understand how human beings function.  However, this is a somewhat unsatisfactory conclusion.  Clearly, what scientific inquiry into human biology is seeking is the translation of the physical representation of a human being into the language representation, and subsequently into a mathematical representation.  
 
Because the discovery of how the brain works amounts to translating the internal neural pathways from a physical representation into a language one, it suggests that language is a kind of ``mirror" into internal neural constructs.  This implies that language, if properly and carefully used, could be used to infer the existence of previously undiscovered neural pathways in the brain.  Furthermore, it could give clues as to how known neural pathways actually function.  

An ability to translate the internal neural pathways from a physical representation into a language one must include the ability to translate the translation mechanism itself.  In principle this should be possible.  To take a biochemical example, the enzymes capable of catalyzing the translation of genetic information from the nucleic acid to the amino acid representation are themselves encoded in the DNA chain, and may themselves be translated.

Ultimately, the scientific exploration of the physical universe, including an understanding of the nature of self-awareness, amounts to a search for self-contained, circular constructs that may be regarded as ``self-replicating" in a sense (i.e. we are looking for tautologies).  These circular constructs are then primitives that are only defined with reference to themselves, and not to any kind of philosophical notion of a higher truth (whether our reality can be described by a finite or infinite number of such ``self-replicating" primitives is an undecidable question).  

To illustrate why this must be the case, it may be argued that any analysis that a self-aware organism may make about itself based on known physical laws is fundamentally circular, because the physical laws being employed were deduced by the brain doing the analysis.  As a result, the physical laws may be regarded as a product of the organismal brain, and so any analysis based on such laws is not really based on any kind of external, ``objective" constructs. 

 As was noted in the discussion on solipsism, from a scientific perspective this is not a problem:  We can take the physical laws as primitives, from which we can develop a set of pathways that lead to the development of self-aware organisms that can deduce those laws.  Alternatively, in the solipsistic point of view, we can take the self-aware organism as a primitive, that then creates a reality with a certain set of physical laws, which allow for the emergence of self-aware organisms.  Both views are equivalent, because both are self-contained.   

\subsection{Using language and set theory to understand brain structure and neural network topology}

We also claim that concepts from logic and set theory, as well as the related field of {\it pattern recognition} \cite{TANNENBAUM, PROSS}, may be highly useful for understanding neural pathway formation.  The reason for this is that the human brain is capable of deducing physical laws, and developing various abstract concepts such as number, infinity, and proof-by-contradiction.  While we may take these constructs as primitives, if we want to fully understand how a collection of neurons forms a human brain, then we must understand how such a collection can develop the various abstract concepts described above.

Set and logic theory provides a foundation for all of mathematics that is based on primitives called {\it sets}, with the derivation of all subsequent concepts via various associations formed between sets.  Any aspect of the observable universe may ultimately be defined using set-theoretic constructs.  Because it is believed that brain function is driven by the formation of neural pathways that define various associations between neurons, it is likely that set theory could be used to infer the structure of neural pathways corresponding to various abstract concepts.  Furthermore, set and logic theory may provide a rigorous framework for understanding the tautologies that emerge in discussions on self-awareness, since the identification of tautologies is one of the central objectives of set theory.

\subsection{Organismal curiousity}

Earlier in this paper, we postulated that organismal curiousity may play a key role for the emergence of the phenomenon of ``awareness of self-awareness".  The argument was based on the claim that, through a process of associative learning, a sufficiently intelligent brain would develop contradictory associations.  If the brain was wired with a natural curiousity, it would attempt to resolve these internal contradictions, which is ultimately impossible because they are tautologies.  The result is that such a brain would be caught in a kind of ``loop", which would lead to the ``pondering" behavior that is characteristic of highly self-aware organisms.

We therefore claim that curiosity is an essential minimal feature that must be incorporated into any machine that is designed to be self-aware.  As a result, it makes sense to conclude this paper with a speculation on the evolutionary basis for the emergence of such behavior in complex organisms.

As an organism with a sufficiently complex brain observes the world, it may develop various associations that provide useful constructs for processing the sensory information it obtains.  If some of the sensory information leads to contradictory associations, one course of action is for the organismal brain to simply ignore the contradictions.  Indeed, holding two contradictory thoughts is a commonly observed phenomenon, termed {\it cognitive dissonance}.  

However, such behavior might not be optimal for organismal survival.  As an example, an organism might be hard-wired to have an innate fear of objects it does not recognize.  This behavior can confer a survival advantage to the organism, since it does not attempt to enter into potentially life-threatening situations that it does not know how to handle.  On the other hand, the disadvantage of such behavior is that it may prevent the organism from taking advantage of new opportunities, such as a previously unseen food source.  Therefore, a more balanced response for an organism is to attempt to obtain more information about its environment, when the incoming sensory information consists of contradictory components and is therefore inconclusive.  

Whatever its evolutionary basis may be, organismal curiousity is a behavior that emerges from a collection of neurons acting under the influence of a reward-punishment-based pathway selection mechanism.  Understanding how such a mechanism, with its associated reward and stress-response-inducing chemicals, leads to the emergence of organismal curiosity, is an important area of both theoretical and experimental work.

\subsection{Associative learning as the basic algorithm for extracting information about a system}

One final issue that we should address is the basis for using associative learning as the algorithm for determining how various systems function.  To formulate the answer to this question, we may note that determining how a system functions is analogous to reading the content of a linear data stream, where the data stream contains the necessary information defining the given system.  In general, if we do not have any a priori information about the system, then we do not know the length of the data stream, or even if the data stream terminates.  Therefore, there is no guarantee that we will be able to fully read the data stream, and hence the question as to whether it will be possible to fully determine how a given system functions is ultimately unresolvable.  

Despite this uncertainty, it is still possible to attempt to read as much of the data stream as possible in a given amount of time, and to attempt as faithful a reconstruction of the system as possible.  The difficulty that arises is that the data stream may be too large to be read at once.  That is, the machine reading the data stream may only be able to read in short strands of data at a time.  Reconstructing the full data stream then requires the machine to look for associations between the various strands.  That is, the machine may notice that one data strand is present with another data strand, so that the two may actually be the successive components of a larger strand within the data stream.  Gradually, as more and more strand-strand associations are determined, the full data stream may be reconstructed.  

Associative learning is therefore a natural tool that arises for reconstructing systems, because it is a mechanism that allows for the integration of partial observations of a system in such a way that the system may be fully decoded and then reconstructed.  It should also be noted that the decoding scheme described here is similar to gene sequencing methods, where many copies of short strands of DNA are analyzed and compared for overlaps, in order to determine the likelihood of various strand-strand associations and thereby reconstruct the whole genome \cite{GENESEQUENCE}.  Gene-gene interactions are also discovered by looking for correlations, or associations, in gene expression data \cite{GENENETWORK}.

\end{document}